\documentclass[aps,twocolumn,showpacs,pre]{revtex4-1}
\usepackage{graphicx,graphics}
\usepackage{amsmath,amssymb,amsfonts}
\usepackage{bbm,bbold}
\usepackage{multirow}
\usepackage{hyperref}
\usepackage{graphicx}

\begin{document}
\title{Random walks on networks with stochastic resetting }
\author{Alejandro P. Riascos$^1$}
\author{Denis Boyer$^1$}
\author{Paul Herringer$^2$}
\author{Jos\'e L. Mateos$^{1,3}$}
\affiliation{${}^1$Instituto de F\'isica,Universidad Nacional Aut\'onoma de M\'exico,\\
Apartado Postal 20-364, 01000 Ciudad de M\'exico, M\'exico}
\affiliation{${}^2$Department of Physics and Astronomy, University of Calgary, \\
Calgary, Alberta T2N 1N4, Canada}
\affiliation{${}^3$Centro de Ciencias de la Complejidad,\\ Universidad Nacional Aut\'onoma de M\'exico, 04510 Ciudad de M\'exico, M\'exico}
\date{\today}
\begin{abstract}
We study random walks with stochastic resetting to the initial position on arbitrary networks. We obtain the stationary probability distribution as well as the mean and global first passage times, which allow us to characterize the effect of resetting on the capacity of a random walker to reach a particular target or to explore a finite network.  We apply the results to rings, Cayley trees, random and complex networks. Our formalism holds for undirected networks and can be implemented from the spectral properties of the random walk without resetting, 
providing a tool to analyze the search efficiency in different structures 
with the small-world property or communities. In this way, we extend the study of resetting processes to the domain of networks.
\end{abstract}


\maketitle
\section{Introduction}
When a stochastic process is occasionally reset, {\it i.e.}, interrupted and restarted from the initial state,
its occupation probability in the configuration space is strongly altered. Interestingly, the mean time needed to reach a given 
target state for the first time can often be minimized with respect to the resetting 
rate \cite{evans2011diffusion,Evans2011JPhysA,reuveni2016optimal,EvansReview2019}. Random search strategies can thus be improved by 
resetting, a fact that finds applications in statistical physics \cite{manrubia1999stochastic,montanari2002optimizing}, 
computer science \cite{luby1993optimal}, enzymatic reactions \cite{reuveni2014role} or 
foraging ecology \cite{boyer2014random,giuggioli2018comparison,Pal2019}. In recent years,
different types of resetting protocols have been considered \cite{pal2016diffusion,nagar2016diffusion,Bhat2016JStat,chechkin2018random} on a variety of underlying processes, such as Brownian motion \cite{evans2011diffusion,Evans2011JPhysA,MajumdarPRE2015}, processes with a drift \cite{montero2013monotonic,ray2019peclet} or models of anomalous diffusion \cite{Kusmierz2014PRL,kusmierz2015optimal,kusmierz2019subdiffusive,maso2019transport}. All these problems are more conveniently studied in relatively simple search spaces, mainly, the semi-infinite line, $\mathbb{R}^D$ \cite{evans2014diffusion}, bounded domains in $1D$ and $2D$ \cite{christou2015diffusion,chatterjee2018diffusion,pal2019first}, or on infinite regular lattices \cite{boyer2014random,FalconPRL2017,Boyer_2019}. 
\\[2mm]
Nevertheless, random walks and related dynamical processes on more complex structures such as networks are at the foundation of statistical 
physics \cite{RevModPhys1976,Hughes,NohRieger2004,FractionalBook2019} and of relevance for a broad range of phenomena and 
applications \cite{NewmanBook,barabasi2016book}. Examples include data science \cite{CoifmanPNAS2005,BlanchardBook2011}, 
synchronization \cite{Arenas2008Synchronization}, epidemic spreading \cite{SatorrasPRL2001,Barter_PRE2016}, 
human mobility \cite{YMoreno_PRE2012,RiascosMateosPlos2017,Barbosa2018}, 
ranking and searching on the web \cite{Brin1998,LeskovecBook2014,ShepelyanskyRevModPhys2015}, 
among others \cite{VespiBook,MasudaPhysRep2017}. In particular, random walks on networks are relevant to the understanding of contact networks between people \cite{RiascosMateosPlos2017}, which is crucial in problems of contact tracing in epidemics such as the current coronavirus disease COVID-19 pandemic \cite{Ferretti2020}.
Whereas lattice random walks have been explored for decades \cite{Polya1921Uber,Montroll1956Random, Montroll1965}, 
the study of local random walks on complex networks is more recent and was introduced by Noh and Rieger \cite{NohRieger2004}. 
Network exploration by random walks is now better understood \cite{NohRieger2004,Tejedor2009PRE,MasudaPhysRep2017}, 
including non-local strategies with long-range hops between distant 
nodes \cite{RiascosMateos2012,RiascosMateosFD2014,Weng2015,Guo2016,Michelitsch2017PhysA,deNigris2017,Estrada2017Multihopper}. 
\\[2mm]
\begin{figure}[!b] 
\begin{center}
\includegraphics*[width=0.5\textwidth]{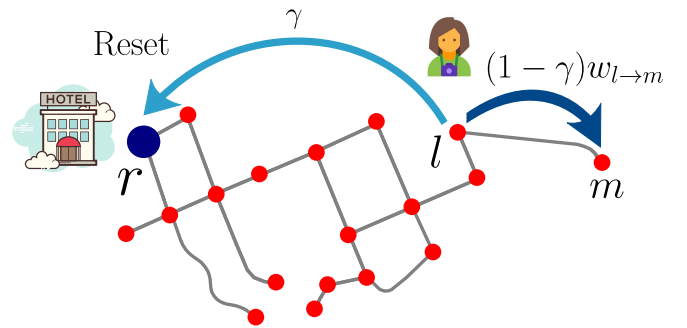}
\end{center}
\vspace{-5mm}
\caption{\label{Figure1} A random walker with resetting can be illustrated as a tourist visiting places in a street network. In the present model, the possible movements from a node $l$ are: a random walk step to an adjacent node (with probability $1-\gamma$), or a relocation to a fixed node $r$ (the hotel) with probability $\gamma$, from which the exploration of the network is resumed.}
\end{figure}
The mean first passage times (MFPT) of random walks subject to resetting have been studied on small graphs or particular social networks \cite{Avrachenkov2014,Avrachenkov2018}, but their properties on arbitrary networks remain little understood, despite their importance. Figure \ref{Figure1} illustrates a dynamics defined by some transition probabilities between adjacent nodes and a resetting probability $\gamma$ to a particular node $r$. Three  important features of many complex and real-world networks are: their finiteness; the small-world effect, characterized by a logarithmic growth of the diameter with the number of nodes \cite{WattsStrogatz1998}; and the presence of communities, {\it i.e.},  subsets of nodes more densely connected to each other than to the other nodes \cite{newman2004finding}. Both the network architecture and the choice of the resetting node should impact significantly the MFPT to a given target node, and more generally, the capacity of the walker to explore the whole network.
\\[2mm]
In this contribution, we develop an extension to arbitrary network topology of the diffusion problem with stochastic resetting of 
\cite{evans2011diffusion,Kusmierz2014PRL}. We deduce general exact expressions for the stationary probability distribution and the first passage times. The analytical results can be expressed in terms of the eigenvalues and eigenvectors of the transition matrix that generates the random walk without resetting.  The methods introduced here can be used to study the effects of resetting on large networks. We apply our findings to regular lattices, trees, random networks and several well-known complex networks. 

\section{General theory}
We study connected (single-component) networks of $N$ nodes labeled by $i=1,\ldots ,N$, and of adjacency matrix $\mathbf{A}$ whose elements are $A_{ij}=A_{ji}=1$ if there is a link between the nodes $i$ and $j$, and $A_{ij}=A_{ji}=0$ otherwise. The links are thus undirected and we also set $A_{ii}=0$ to avoid self-loops. The degree of the node $i$ is denoted as  $k_i=\sum_{l=1}^N A_{il}$. On this structure, we consider a random walker in discrete time and starting at $t=0$ from $i$. The walker performs at $t=1,2,\ldots$ two types of steps: 1) a jump to one of the neighbors of the node currently occupied (all neighbors being equiprobable), and 2) a resetting to a fixed node $r$. Actions 1) and 2) occur with probability $1-\gamma$ and $\gamma$, respectively. 

\subsection{Occupation probability}

Without resetting ($\gamma=0$), the probability to hop to $m$ from $l$ is $w_{l\to m}= A_{lm}/k_l$. This random walk is described by the transition matrix $\mathbf{W}$ with elements $w_{l\to m}$ for $l,m=1,\ldots,N$ \cite{NohRieger2004}. With the incorporation of resetting, the occupation probability follows the master equation
\begin{equation}
 \label{mastereq1}
 P_{ij}(t+1;r,\gamma) = (1-\gamma)\sum_{l=1}^N P_{il}(t;r,\gamma)w_{l\to j}+\gamma\delta_{rj},
\end{equation}
where $P_{ij}(t;r,\gamma)$ denotes the probability to find the walker at $j$ at time $t$, given the initial position $i$, resetting node $r$ and resetting probability $\gamma$ 
($\delta_{rj}$ denotes the Kronecker delta).
The first term on the right-hand side of Eq. (\ref{mastereq1}) represents hops between adjacent nodes whereas
the second term describes resetting to $r$. 
Let us define the transition probability matrix $\mathbf{\Pi}(r,\gamma)$ with elements $
\pi_{l \to m}(r,\gamma)\equiv (1-\gamma) w_{l\to m}+\gamma\,\delta_{rm}$. Eq. (\ref{mastereq1}) takes the simpler form of a Markov chain 
\begin{equation}\label{mastermarkov}
P_{ij}(t+1;r,\gamma) = \sum_{l=1}^N  P_{il} (t;r,\gamma) \pi_{l\to j}(r,\gamma), 
\end{equation}
where $\sum_{m=1}^N \pi_{l \to m}(r,\gamma)=1$. The matrix 
$\mathbf{\Pi}(r,\gamma)$ completely entails the dynamics with resetting. As we are considering connected undirected networks, the process defined by Eq. (\ref{mastermarkov}) is able to reach all the nodes of the network if the resetting probability $\gamma$ is $<1$. Like $\mathbf{W}$,  $\mathbf{\Pi}(r,\gamma)$ is a stochastic matrix:  knowing its eigenvalues and eigenvectors allows the calculation of the occupation probability at any time, including the stationary distribution, as well as the mean first passage time to any node. 

We first analyze how the eigenvalues and eigenvectors of $\mathbf{\Pi}(r,\gamma)$  are related to those of $\mathbf{W}$, which is recovered in the limit $\gamma=0$ and can be readily computed numerically or analytically in some cases. Employing Dirac's notation,
we denote the eigenvalues of the matrix $\mathbf{W}$, which is not symmetric in general, as $\lambda_l$ (where $\lambda_1=1$), and its right and left eigenvectors as $\left|\phi_l\right\rangle$ and $\left\langle\bar{\phi}_l\right|$, respectively, for $l=1,2,\ldots,N$. Similarly, the eigenvalues of $\mathbf{\Pi}(r,\gamma)$ are denoted as $\zeta_l(r,\gamma)$ and its eigenvectors as $\left|\psi_l(r,\gamma)\right\rangle$ and $\left\langle\bar{\psi}_l(r,\gamma)\right|$. 
\\[2mm]
Let us analyze the connection between the eigenvalues $\lambda_l$ and $\zeta_l(r,\gamma)$. We may use the identity  
\begin{equation}
\mathbf{\Pi}(r,\gamma)=(1-\gamma)\mathbf{W}+\gamma \mathbf{\Theta}(r), 
\end{equation}
where the elements of the matrix $\mathbf{\Theta}(r)$ are $\Theta_{lm}(r)=\delta_{mr}$. Namely, $\mathbf{\Theta}(r)$ has entries $1$ in the $r^{th}$-column and null entries everywhere else. We obtain (see Appendix \ref{AppendixA} for details)
\begin{equation}\label{eigvals_zeta}
\zeta_l(r,\gamma)=
\begin{cases}
1 \qquad &\mathrm{for}\qquad l=1,\\
(1-\gamma)\lambda_l \qquad &\mathrm{for}\qquad l=2,3,\ldots, N.
\end{cases}
\end{equation}
This result reveals that the eigenvalues are independent of the choice of the resetting node $r$. The left eigenvectors of $\mathbf{\Pi}(r,\gamma)$ are further given by (see also Appendix \ref{AppendixA} for details)
\begin{equation}\label{psil1}
\left\langle\bar{\psi}_1(r,\gamma)\right|=\left\langle\bar{\phi}_1\right|
+\sum_{m=2}^N\frac{\gamma}{1-(1-\gamma)\lambda_m}\frac{\left\langle r|\phi_m\right\rangle}{\left\langle r|\phi_1\right\rangle}\left\langle\bar{\phi}_m\right|
\end{equation}
whereas $\left\langle\bar{\psi}_l(r,\gamma)\right|=\left\langle\bar{\phi}_l\right|$ for $l=2,\ldots,N$. Similarly, the right eigenvectors are given by:
$\left|\psi_1(r,\gamma)\right\rangle=\left|\phi_1\right\rangle$ and
\begin{equation}
\left|\psi_l(r,\gamma)\right\rangle=
\left|\phi_l\right\rangle-\frac{\gamma}{1-(1-\gamma)\lambda_l}\frac{\left\langle r|\phi_l\right\rangle }{\left\langle r|\phi_1\right\rangle}\left|\phi_1\right\rangle,
\end{equation}
for $l=2,\ldots,N$, where $|r\rangle$ denotes the vector with all its components equal to 0 except the $r$-th one, which is equal to 1. With the left and right eigenvectors at hand, one can use the spectral representation $\mathbf{\Pi}(r,\gamma)=\sum_{l=1}^N\zeta_l(r,\gamma)\left|\psi_l(r,\gamma)\right\rangle\left\langle\bar{\psi}_l(r,\gamma)\right|$. 
\\[2mm]
This spectral approach for discrete time random walks is also valid for continuous-time random walks (this case is analyzed in Appendix \ref{AppendixB}). At unit hopping rate, the dynamics is defined by the modified Laplacian $\hat{\mathcal{L}}(r,\gamma)=\mathbb{1}-\mathbf{\Pi}(r,\gamma)$ ($\mathbb{1}$ denotes the $N\times N$ identity matrix) which has the same eigenvectors  of $\mathbf{\Pi}(r,\gamma)$ and eigenvalues $\xi_m(r,\gamma)=1-\zeta_m(r,\gamma)$. Our findings for the spectral properties of $\hat{\mathcal{L}}(r,\gamma)$ coincide with the general approach of \cite{Touchette_PRE2018} in the context of classical and quantum transport with resetting.
\\[2mm]
In the discrete case, the occupation
probability of the process described by Eq. (\ref{mastermarkov}) is given by
\begin{equation}
P_{ij}(t;r,\gamma)=\left\langle i|\mathbf{\Pi}(r,\gamma)^t|j\right\rangle. 
\end{equation}
We deduce
\begin{multline}\label{Pijpsivectors}
P_{ij}(t;r,\gamma)=\left\langle i\left|\psi_1(r,\gamma)\right\rangle \left\langle\bar{\psi}_1(r,\gamma)\right|j\right\rangle\\
+\sum_{l=2}^N[(1-\gamma)\lambda_l]^t\left\langle i|\psi_l(r,\gamma)\right\rangle\left\langle\bar{\psi}_l(r,\gamma)|j\right\rangle.
\end{multline}
The first term in Eq. (\ref{Pijpsivectors}) defines the long time, stationary distribution $P_j^\infty(r,\gamma)=\left\langle i\left|\psi_1(r,\gamma)\right\rangle \left\langle\bar{\psi}_1(r,\gamma)\right|j\right\rangle$. By using  Eq. (\ref{psil1}) and $\left|\psi_1(r,\gamma)\right\rangle=\left|\phi_1\right\rangle$, we obtain 
\begin{equation}\label{Pinfvectors}
P_j^\infty(r,\gamma)=\frac{k_j}{\sum_{m=1}^N k_m}+\gamma\sum_{l=2}^N\frac{\left\langle r|\phi_l\right\rangle \left\langle\bar{\phi}_l|j\right\rangle}{1-(1-\gamma)\lambda_l},
\end{equation}
where we have used the identity $\left\langle i|\phi_1\right\rangle \left\langle\bar{\phi}_1|j\right\rangle=\frac{k_j}{\sum_{m=1}^N k_m}$ for the equilibrium distribution of the usual random walk on networks
\cite{NohRieger2004,MasudaPhysRep2017}. The second term of $P_j^\infty(r,\gamma)$ in Eq. (\ref{Pinfvectors}) corresponds to a non-equilibrium part, which is a consequence of the resetting dynamics \cite{evans2011diffusion}.  One obtains the occupation probability in terms of the spectral properties of $\mathbf{W}$ 
\begin{multline}\label{Pijspect}
P_{ij}(t;r,\gamma)=P_j^\infty(r,\gamma)\\
+\sum_{l=2}^N(1-\gamma)^t\lambda_l^t\left[\left\langle i|\phi_l\right\rangle \left\langle\bar{\phi}_l|j\right\rangle-\gamma\frac{\left\langle r|\phi_l\right\rangle \left\langle\bar{\phi}_l|j\right\rangle}{1-(1-\gamma)\lambda_l} \right].
\end{multline}

\subsection{Mean first passage and return times}

The expression for the MFPT to the target $j$ starting from $i$ 
can be deduced from the general convolution property with $P_{ij}(t;r,\gamma)$ for Markov processes, see Appendix \ref{sec:mfptApp} or \cite{Hughes}. It is given by  
\begin{equation}
\left\langle T_{ij}(r,\gamma)\right\rangle=\frac{1}{P_j^\infty(r,\gamma)}\left[\delta_{ij}+R_{jj}^{(0)}(r,\gamma)-R_{ij}^{(0)}(r,\gamma)\right], 
\end{equation}
where 
\begin{equation}
R_{ij}^{(0)}(r,\gamma)\equiv \sum_{t=0}^\infty [P_{ij}(t;r,\gamma)-P_j^\infty(r,\gamma)].
\end{equation}
Using Eq. (\ref{Pijspect}), one obtains in the case of resetting to the origin, {\it i.e.}, $r=i$ (see Appendix \ref{sec:mfptApp}):
\begin{multline}\label{MFPT_reset}
\left\langle T_{ij}(\gamma)\right\rangle=\frac{\delta_{ij}}{P_j^\infty(i,\gamma)}\\
+
\frac{1}{P_j^\infty(i,\gamma)}\sum_{l=2}^N\frac{
\left\langle j|\phi_l\right\rangle \left\langle\bar{\phi}_l|j\right\rangle-\left\langle i|\phi_l\right\rangle \left\langle\bar{\phi}_l|j\right\rangle
}{1-(1-\gamma)\lambda_l}.
\end{multline}
Note that the case $j=i$ corresponds to the {\it mean first return time} to $i$, hence $\left\langle T_{ii}(\gamma)\right\rangle$ is non-zero but equal to $1/P_i^\infty(i,\gamma)$, in agreement with Kac's lemma on mean recurrence times \cite{kac1947notion}.
It is also useful to quantify the ability of a process to explore the whole network.
To this purpose,
we define $\mathrm{T}(i,\gamma)$ as the global MFPT starting from node $i$,
\begin{equation}\label{globalMFPT}
\mathrm{T}(i,\gamma)\equiv\frac{1}{N}\sum_{j=1}^N \langle T_{ij}(\gamma)\rangle.
\end{equation}
The results in Eqs. (\ref{eigvals_zeta})-(\ref{MFPT_reset}) apply to random walks with resetting on any finite, connected and undirected network. The eigenvalues and eigenvectors of $\mathbf{W}$ may be obtained by direct numerical calculation, or analytically in particular cases. We next explore the effects of resetting in different network topologies. 

\section{Rings}
\begin{figure*}[!t] 
\begin{center}
\includegraphics*[width=0.98\textwidth]{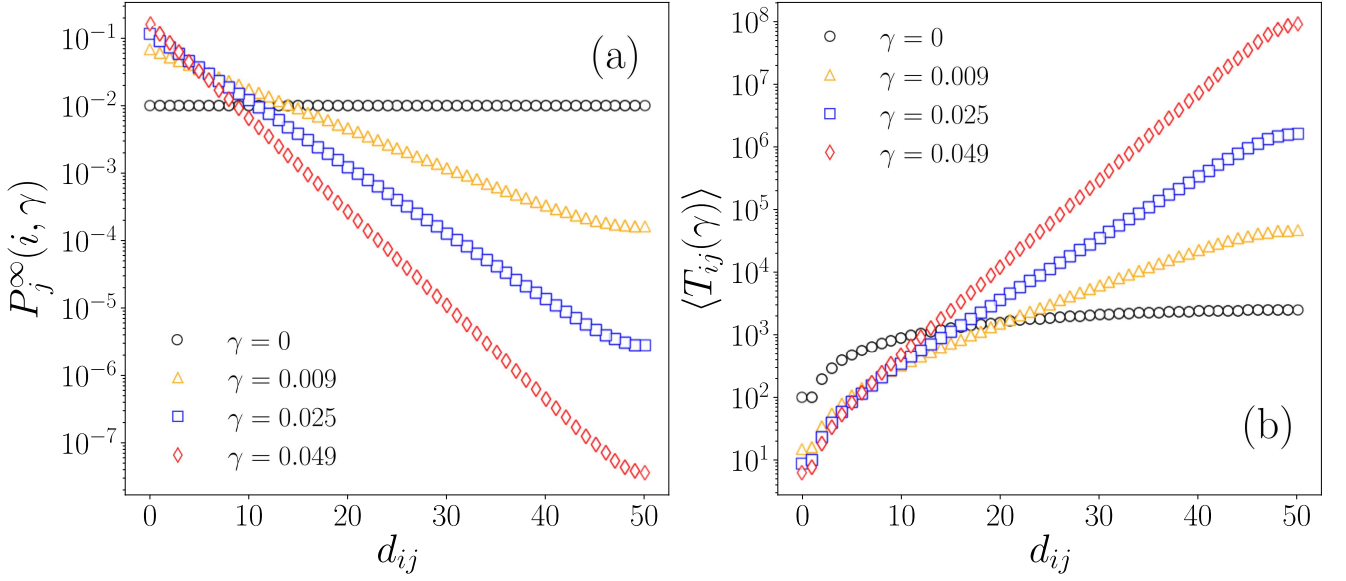}
\end{center}
\vspace{-5mm}
\caption{\label{Figure2} Stationary distribution and MFPT for random walks with resetting on a ring with $N=100$ nodes. (a)  $P_j^\infty(i,\gamma)$ as given by Eq. (\ref{Pinf_ring})  and (b) $\left\langle T_{ij}(\gamma)\right\rangle$ as given by Eq. (\ref{MFPT_ring}), as a function of the distance $d_{ij}$ between the initial node $i$ and the target node $j$ for different values of $\gamma$.}
\end{figure*}
We start our discussion with the analysis of the finite ring, {\it i.e}, the one-dimensional lattice with periodic boundary condition. In this case, $\mathbf{W}$ is a circulant matrix \cite{VanMieghem2011,RiascosMateosFD2015} with eigenvalues $\lambda_l=\cos\left[\frac{2\pi(l-1)}{N}\right]$ and eigenvectors with components $\langle i|\phi_l\rangle=\frac{1}{\sqrt{N}}e^{-\mathrm{i}\frac{2\pi(l-1)(i-1)}{N}}$ and $\langle \bar{\phi}_l|j\rangle=\frac{1}{\sqrt{N}}e^{\mathrm{i}\frac{2\pi(l-1)(j-1)}{N}}$ (here $\mathrm{i}\equiv\sqrt{-1}$) for $l=1,\ldots,N$. The stationary distribution (\ref{Pinfvectors}) for a ring takes the form
\begin{align}\nonumber
P_j^\infty(i,\gamma)&=\frac{1}{N}+\gamma\sum_{l=2}^N\frac{\left\langle i|\phi_l\right\rangle \left\langle\bar{\phi}_l|j\right\rangle}{1-(1-\gamma)\lambda_l}\\\nonumber
&=\frac{1}{N}+\frac{\gamma}{N}\sum_{l=2}^N\frac{e^{-\mathrm{i}\frac{2\pi(l-1)(i-j)}{N}}}{1-(1-\gamma)\cos\left[\frac{2\pi(l-1)}{N}\right]}\\
&=\frac{1}{N}+\frac{\gamma}{N}\sum_{l=2}^N \frac{\cos\left(\varphi_l\,d_{ij}\right)}{1-(1-\gamma)\cos(\varphi_l)}
\label{Pinf_ring}
\end{align}
with $\varphi_l\equiv\frac{2\pi}{N}(l-1)$, and where $d_{ij}$ is the distance between $i$ and $j$ [note that $\cos\left[\varphi_l(i-j)\right]=\cos\left(\varphi_l\,d_{ij}\right)$, see also \cite{RiascosMateosFD2014}]. For the MFPT, 
Eq. (\ref{MFPT_reset}) is recast as
\begin{align}\nonumber
&\left\langle T_{ij}(\gamma)\right\rangle=\frac{\delta_{ij}+
\sum\limits_{l=2}^N\frac{\left\langle j|\phi_l\right\rangle \left\langle\bar{\phi}_l|j\right\rangle-\left\langle i|\phi_l\right\rangle \left\langle\bar{\phi}_l|j\right\rangle
}{1-(1-\gamma)\lambda_l}}{P_j^\infty(i,\gamma)}\\\nonumber
&\qquad\qquad=\frac{\delta_{ij}+\frac{1}{N}
\sum\limits_{l=2}^N\frac{1-e^{-\mathrm{i}\frac{2\pi(l-1)(i-j)}{N}}}{1-(1-\gamma)\cos\left[\frac{2\pi(l-1)}{N}\right]}}{P_j^\infty(i,\gamma)}\\
&=\frac{1}{P_j^\infty(i,\gamma)}\left[\delta_{ij}+\frac{1}{N}
\sum_{l=2}^N\frac{1-\cos\left(\varphi_l\,d_{ij}\right)}{1-(1-\gamma)\cos(\varphi_l)}\right]. \label{MFPT_ring}
\end{align}
Figure \ref{Figure2} displays the analytical expressions in Eqs. (\ref{Pinf_ring}) and (\ref{MFPT_ring}) for $N=100$, as a function of the distance $d_{ij}$. These quantities exhibit exponential behaviors.
\\[2mm]
In the limit $N\to\infty$,  we recover the infinite one-dimensional lattice, where  $\varphi=\frac{2\pi}{N}(l-1)$ can be considered as a continuous variable with $d\varphi=\frac{2\pi}{N}$. The stationary distribution $P_j^\infty(i,\gamma)$ in Eq. (\ref{Pinf_ring}) for the infinite ring takes the form 
\begin{equation}\label{Pinf_infring}
P_j^\infty(i,\gamma)=\frac{\gamma}{2\pi}\int_0^{2\pi} \frac{\cos( d_{ij}\,\varphi)}{1-(1-\gamma)\cos(\varphi)}d\varphi.
\end{equation}
To evaluate Eq. (\ref{Pinf_infring}), we define the integral 
\begin{equation}
\mathcal{I}(x,b)=\frac{1}{2\pi}\int_0^{2\pi} \frac{\cos(x\theta)}{1-b\cos(\theta)}d\theta, \,\,0\leq b<1,\, x\geq 0
\end{equation}
and, by using $b=\frac{2a}{1+a^2}$, we have
\begin{multline}
\mathcal{I}(x,b)=\frac{a^2+1}{2\pi}\int_0^{2\pi} \frac{\cos(x\theta)}{1+a^2-2a\cos(\theta)}d\theta\\
=\frac{a^2+1}{(a^2-1)a^x}\, , \qquad \text{for}\qquad a^2>1.
\end{multline}
where we have used the identity $\frac{1}{2\pi}\int_0^{2\pi} \frac{\cos(x\theta)}{1+a^2-2a\cos(\theta)}d\theta=\frac{1}{(a^2-1)a^x}$  (see, {\it e.g.}, \cite{Boyer_2019}). Hence, using $a=\frac{1}{b}+\sqrt{\frac{1}{b^2}-1}$
\begin{equation}
\frac{1}{2\pi}\int_0^{2\pi} \frac{\cos(x\theta)}{1-b\cos(\theta)}d\theta=\frac{\left(\frac{1+\sqrt{1-b^2}}{b}\right)^{-x}}{\sqrt{1-b^2}}.
\end{equation}
Combining this result with $b=1-\gamma$ in Eq. (\ref{Pinf_infring}), we obtain
\begin{equation}\label{Pinf_infring2}
P_j^\infty(i,\gamma)=\sqrt{\frac{\gamma}{2-\gamma} }\left(\frac{\sqrt{(2-\gamma ) \gamma }+1}{1-\gamma}\right)^{-d_{ij}}. 
\end{equation}
In the limit of small resetting probability $0<\gamma\ll 1$, $\sqrt{\frac{\gamma}{2-\gamma} }=\frac{\sqrt{2\gamma }}{2}+O\left(\gamma ^{3/2}\right)$ and $\log\left(\frac{\sqrt{(2-\gamma ) \gamma }+1}{1-\gamma}\right)=\sqrt{2\gamma }+O\left(\gamma ^{3/2}\right)$. Consequently, the stationary distribution satisfies
\begin{equation}
P_j^\infty (i,\gamma)\approx\frac{\sqrt{2\gamma }}{2}e^{-\sqrt{2\gamma }d_{ij}}\qquad \text{for}\qquad 0<\gamma\ll 1,
\end{equation}
which coincides with the exponential non-equilibrium steady state of the one-dimensional Brownian motion with diffusion coefficient $1/2$ and resetting rate $\gamma$ \cite{evans2011diffusion}.
\\[4mm]
We now specify our results on the MFPT for an infinite ring. In the case $N\to \infty$, Eq. (\ref{MFPT_ring}) takes the form
\begin{equation}
\left\langle T_{ij}(\gamma)\right\rangle=\frac{\delta_{ij}+\frac{1}{2\pi}\int_0^{2\pi}\frac{1-\cos( d_{ij}\,\varphi)}{1-(1-\gamma)\cos(\varphi)}d\varphi}{P_j^\infty(i,\gamma)}
\end{equation}
with $P_j^\infty(i,\gamma)$ given by Eq. (\ref{Pinf_infring2}). In particular, if $i=j$, we obtain the mean first return time to the starting/resetting point
\begin{equation}\label{MFPTii_infring}
\left\langle T_{ii}(\gamma)\right\rangle=\frac{1}{P_i^\infty(i,\gamma)}=\sqrt{\frac{2-\gamma}{\gamma} } .
\end{equation}
On the other hand, if $i\neq j$
\begin{align}\nonumber
&\left\langle T_{ij}(\gamma)\right\rangle=\frac{1}{P_j^\infty(i,\gamma)}\frac{1}{2\pi}\int_0^{2\pi}\frac{1-\cos( d_{ij}\,\varphi)}{1-(1-\gamma)\cos(\varphi)}d\varphi\\
&=\frac{1}{P_j^\infty(i,\gamma)}\frac{1}{2\pi}\int_0^{2\pi}\frac{1}{1-(1-\gamma)\cos(\varphi)}d\varphi-\frac{1}{\gamma}.
\end{align}
Using the identity $\frac{1}{2\pi}\int_0^{2\pi} \frac{1}{1-(1-\gamma)\cos(\varphi)}d\varphi=\frac{1}{\gamma}\sqrt{\frac{\gamma}{2-\gamma} }$,
\begin{equation}\label{MFPTij_infring}
\left\langle T_{ij}(\gamma)\right\rangle=\frac{1}{\gamma}\left(\frac{\sqrt{(2-\gamma ) \gamma }+1}{1-\gamma}\right)^{d_{ij}}-\frac{1}{\gamma}  \qquad \text{for}\qquad i\neq j. 
\end{equation}
Combining the results in Eqs. (\ref{MFPTii_infring}) and (\ref{MFPTij_infring}) gives
\begin{equation}\label{MFPT_infring}
\left\langle T_{ij}(\gamma)\right\rangle=
\begin{cases}
\sqrt{\frac{2-\gamma}{\gamma}} \qquad & j=i,\\
\frac{1}{\gamma}\left[
\left(\frac{\sqrt{(2-\gamma ) \gamma }+1}{1-\gamma}\right)^{d_{ij}}-1\right] & j\neq i.
\end{cases}
\end{equation}
In particular, in the limit of small resetting $\gamma\ll 1$ and $d_{ij}>0$, on obtains $\langle T_{ij}(\gamma)\rangle \approx \frac{1}{\gamma}\left[e^{\sqrt{2\gamma} d_{ij}}-1\right]$, which is non-monotonic with $\gamma$. Solving $\partial \langle T_{ij}(\gamma)\rangle/\partial \gamma=0$ we deduce the value of $\gamma$, $\gamma^*\simeq 1.26982/d_{ij}^2$, that minimizes the MFPT to a target at distance $d_{ij}\gg 1$. These results also coincide with those of the continuous limit \cite{evans2011diffusion}.
\begin{figure*}[!t] 
\begin{center}
\includegraphics*[width=0.98\textwidth]{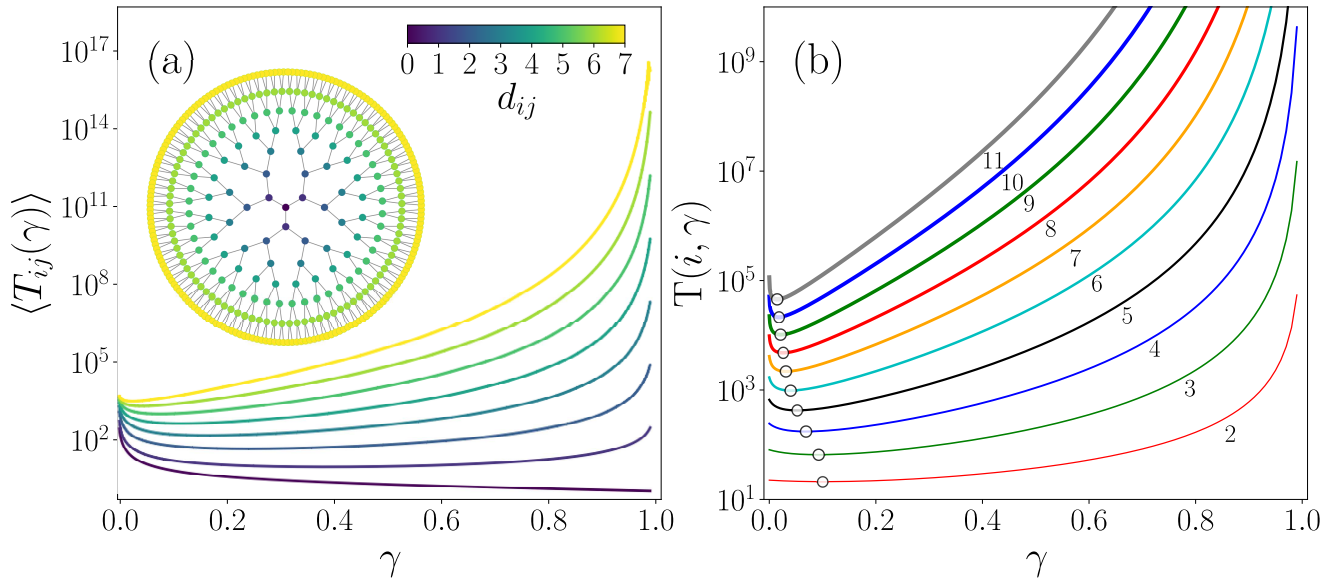}
\end{center}
\vspace{-5mm}
\caption{\label{Figure3} Random walks with stochastic resetting to the central node on Cayley trees with coordination number $z=3$ and $n$ shells. (a) MFPT $\langle T_{ij}(\gamma)\rangle$ vs. $\gamma$ in a Cayley tree with $n=7$ shells ($N=382$ nodes). The results are presented as a family of curves defined by the distance $d_{ij}$ (shown in the color bar) between the central node $i$ and the target node $j$. (b) Global MFPT $\mathrm{T}(i,\gamma)$ defined in Eq. (\ref{globalMFPT}) with $0\leq\gamma\leq 0.99$ for different Cayley trees with $n$ shells. In each curve we include the number $n$ and the circles indicate the minima for Cayley trees with different numbers of shells.}
\end{figure*}

\section{Cayley trees} 
We now consider finite Cayley trees of coordination number $z$ and composed of $n$ shells (see Fig. \ref{Figure3}). The nodes of the last shell have degree $1$, whereas the other nodes have degree $z$.
We display the MFPT $\langle T_{ij}(\gamma)\rangle$ as a function of $\gamma$ in Fig. \ref{Figure3}(a), where $n=7$ and $z=3$ ($N=382$). The starting and resetting position $i$ is the central node. Keeping the distance $d_{ij}(=0,1,\ldots ,n)$ between $i$ and the target $j$ fixed,
we see how resetting modifies the MFPT in comparison with the normal random walk ($\gamma=0$). The mean first return time $\langle T_{ii}(\gamma)\rangle$ (or $d_{ij}=0$) decreases monotonically with $\gamma$, whereas for each positive distance there is a value $\gamma^{\star}$ for which  $\langle T_{ij}(\gamma^*)\rangle$ is minimum, namely, that optimizes the capacity to reach a target at distance $d_{ij}$. Figure \ref{Figure3}(b) displays a similar behavior for the global time $\mathrm{T}(i,\gamma)$, see Eq. (\ref{globalMFPT}), in several Cayley trees of varying $n$. Clearly, $\gamma^*$ decreases with $n$.
\\[2mm]
The limit $n\to \infty$ can be solved analytically by using a general relation between the survival probabilities of discrete time processes with and without resetting \cite{Kusmierz2014PRL}. We recall some basic results on the first passage properties of simple random walks on Cayley trees, see {\it e.g.} \cite{hughes1982random,cassi1989random}, as a preliminary step to further incorporate resetting. Let us consider an infinite Cayley tree with coordination number $z$, a random walk initially at the origin node $0$, and a target node at a distance $d$. We define the survival probability $Q^{(0)}_d(t)$ as the probability that the walker has not reached the target site after $t$ steps, in the absence of resetting. Owing to translational invariance, we write the \lq\lq backward" equation
\begin{equation}\label{backQ}
    Q^{(0)}_d(t)=\frac{z-1}{z}Q^{(0)}_{d+1}(t-1)
    +\frac{1}{z}Q^{(0)}_{d-1}(t-1),
\end{equation}
which asserts that, after the first step (thus with $t-1$ steps to go), with probability $1/z$, the walker can be one unit closer to the target, or with probability $(z-1)/z$, one unit further away. The boundary and initial conditions are
\begin{equation}\label{condQ}
    Q^{(0)}_0(t)=0\quad {\rm and} \quad Q^{(0)}_{d>0}(t=0)=1.
\end{equation}
We introduce the discrete Laplace transform $\widetilde{Q}^{(0)}_d(s)=\sum_{t=0}^{\infty} s^tQ^{(0)}_d(t)$, which from Eq. (\ref{condQ}) must satisfy $\widetilde{Q}^{(0)}_{0}(s)=0$ and $\widetilde{Q}^{(0)}_{d>0}(s=0)=1$. Transforming equation (\ref{backQ}) gives, for $d>0$
\begin{equation}
    \widetilde{Q}^{(0)}_d(s)=1+s\frac{z-1}{z}\widetilde{Q}^{(0)}_{d+1}(s)+\frac{s}{z}\widetilde{Q}^{(0)}_{d-1}(s).
\end{equation}
We look for solutions of the form 
$\widetilde{Q}^{(0)}_d(s)=a+Y_d$. By substitution we find $a=1/(1-s)$ and that $Y_d$ obeys the recursion relation
\begin{equation}
    s\frac{z-1}{z}Y_{d+1}-Y_d+\frac{s}{z}Y_{d-1}=0,
\end{equation}
which is easily solved as $Y_{d}=C_1\nu_1^{d}+C_2\nu_2^{d}$, with 
\begin{align}\nonumber
    \nu_{1}(s)&=
    \frac{z-\sqrt{z^2-4(z-1)s^2}}{2s(z-1)},\\\label{nu}
    \nu_{2}(s)&=
    \frac{z+\sqrt{z^2-4(z-1)s^2}}{2s(z-1)},
\end{align}
and $C_1$, $C_2$ two constants. From $Q_d^{(0)}(t=0)=1$,
the condition $\lim_{s\rightarrow 0}\widetilde{Q}^{(0)}_d(s)\rightarrow1$ must be fulfilled for all $d>0$. Whereas $\nu_1\simeq s/z\rightarrow 0$ at small $s$, $\nu_2\simeq \frac{z}{s(z-1)}\rightarrow\infty$, which imposes $C_2=0$. The second condition $\widetilde{Q}^{(0)}_{0}(s)=0$ is enforced by choosing $C_1=-1/(1-s)$. We deduce
\begin{equation}\label{Qtildenoreset}
    \widetilde{Q}^{(0)}_d(s)=
    \frac{1-[\nu_1(s)]^d}{1-s}.
\end{equation}
The large time behavior of $Q^{(0)}_d(t)$ is deduced from that of $\widetilde{Q}^{(0)}_d(s)$ as $s\rightarrow 1$. Noting that $1/(1-s)$ is the Laplace transform of 1 and that $\lim_{s\rightarrow 1}\nu_1(s)<1$, we deduce from Eq. (\ref{Qtildenoreset}) that $\lim_{t\rightarrow\infty} Q^{(0)}_d(t)=1-\nu_1^d(s=1)$, or
\begin{equation}
    Q^{(0)}_d(t)\rightarrow 1-(1-z)^{-d}\quad{\rm as}\quad t\rightarrow\infty.
\end{equation}
Hence, the probability that the walker ever reaches the target is $(z-1)^{-d}$ \cite{hughes1982random}. The MFPT is readily deduced from the general relation $\langle T_d\rangle=\sum_{t=0}^{\infty} Q_d(t)=\widetilde{Q}_d(s=1)$, which, from Eq. (\ref{Qtildenoreset}), is infinite.

When resetting is present, we can use the renewal approach exposed in \cite{kusmierz2014first}, allowing to derive the survival probability in the Laplace domain, $\widetilde{Q}_d(s)$, as a function of this quantity in the absence of resetting, $\widetilde{Q}^{(0)}_d(s)$. One notices that ${Q}_d(t)$ can be decomposed as the sum of two contributions: $(i)$ either the walker has never reset since $t=0$, which happens with probability $(1-\gamma)^{t}$, $(ii)$ or the last reset happened at a time $1\le\tau\le t$, an eventuality that occurs with probability $\gamma(1-\gamma)^{t-\tau}$. One obtains \cite{kusmierz2014first}
\begin{multline}\label{renewal}
    Q_d(t)=(1-\gamma)^t Q^{(0)}_d(t)\\
+\sum_{\tau=1}^t\gamma(1-\gamma)^{t-\tau}
    Q_d(\tau-1)Q^{(0)}_d(t-\tau).
\end{multline}
The second term asserts that the walker has survived the first $\tau-1$ time steps following the dynamics with resetting, as well as the last $t-\tau$ steps following the reset-free process. Taking the discrete Laplace transform of Eq. (\ref{renewal}) gives
\begin{equation}\label{Qtilde}
    \widetilde{Q}_d(s)=\frac{\widetilde{Q}^{(0)}_d(s(1-\gamma))}
    {1-\gamma s \widetilde{Q}^{(0)}_d(s(1-\gamma))}.
\end{equation}
The MFPT $\langle T_d\rangle=\widetilde{Q}_d(s=1)$ is deduced from Eq. (\ref{Qtilde}) and Eq. (\ref{Qtildenoreset})
\begin{equation}
   \langle T_d\rangle=\frac{1}{\gamma}\left[\nu_1(1-\gamma)^{-d}-1\right], 
\end{equation}
which is a finite quantity. Hence, with Eq. (\ref{nu}), the complete expression is
\begin{equation}\label{mfptcayley} 
    \langle T_d\rangle=\frac{1}{\gamma}\left[ 
\left(\frac{2(1-\gamma)(z-1)}{z-\sqrt{z^2-4(1-\gamma)^2(z-1)}} \right)^d-1
    \right].
\end{equation}
\begin{figure*}[!t] 
\begin{center}
\includegraphics*[width=1.0\textwidth]{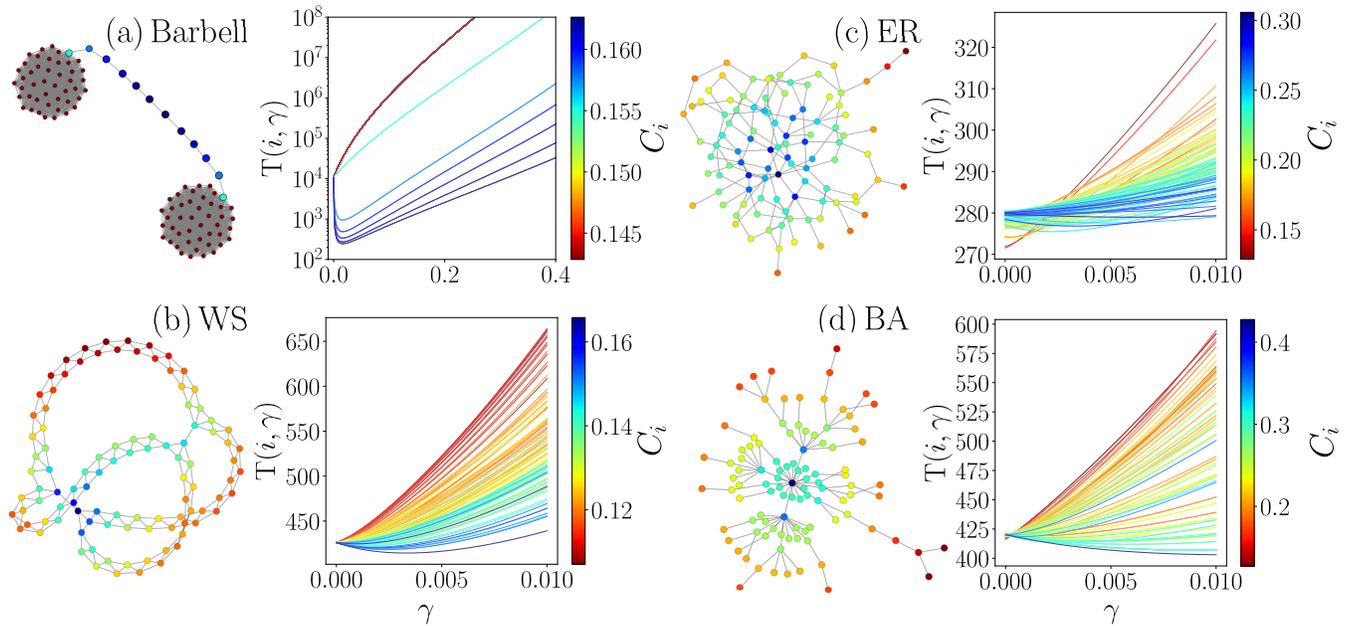}
\end{center}
\vspace{-3mm}
\caption{\label{Figure4} 
Global time in networks with $N=100$ nodes: (a) Barbell; (b) Watts-Strogatz; (c) Erd\"{o}s-R\'enyi and; (d) Barab\'asi-Albert, where each newly introduced node connects to $m$ previous nodes ($m=1$). We depict the global time $\mathrm{T}(i,\gamma)$ as a function of $\gamma$ for all the nodes $i=1,\ldots,N$. To identify the effects of resetting, we colored each node $i$ and its corresponding curve according to its closeness centrality $C_i \equiv \frac{N}{\sum_{j=1}^N d_{ij}}$. }
\end{figure*}
It is easy to check that $\langle T_d\rangle\simeq [\nu_1(1)^{-d}-1]/\gamma\rightarrow\infty$ as $\gamma\rightarrow0$,
and that $\langle T_d\rangle\simeq z^d/(1-\gamma)^{d}\rightarrow\infty$ as $\gamma\rightarrow 1$. Thus $\langle T_d\rangle$ has a minimum for some optimal value $\gamma^*$. The optimal resetting probability is obtained from solving $\partial  \langle T_d\rangle/\partial \gamma=0$, or
\begin{equation}\label{eqopt}
    1-\nu_1(1-\gamma^*)^{d}=d\gamma^* \frac{\nu_1'(1-\gamma^*) }{\nu_1(1-\gamma^*)}.
\end{equation}
In the limit $d\gg1$, one can neglect $\nu_1(1-\gamma^*)^{d}$ in Eq. (\ref{eqopt}) to obtain
\begin{equation}\label{gammaopt}
    \gamma^*\simeq \frac{1}{d}\left(\frac{z-2}{z}\right).
\end{equation}
Hence the optimal resetting rate tends to 0 at large $d$ differently than on regular lattices, where $\gamma^*\sim 1/d^2$ \cite{evans2011diffusion}. This is due to the fact that random walks on Cayley trees are effectively drifting away from their starting point \cite{cassi1989random} and thus travel a distance $d$ during a time of order $d$, instead of $d^2$. The MFPT at optimality is readily obtained by substituting Eq. (\ref{gammaopt}) into (\ref{mfptcayley})
\begin{equation}\label{mfptopt}
    \langle T_d^*\rangle\simeq d \frac{z(z-1)^d}{z-2}.
\end{equation}
This result can be interpreted as follows. The quantity $\frac{z(z-1)^d}{z-2}$ in Eq. (\ref{mfptopt}) represents the total number of nodes located at a distance $d$ or smaller from the origin, that we denote as $N_d$. It stems from the equality $N_d=1+z\sum_{k=0}^{d-1}(z-1)^k\simeq \frac{z(z-1)^d}{z-2}$ at large $d$. We deduce $d\simeq\frac{\ln N_d}{\ln(z-1)}$ and
\begin{equation}
     \langle T_d^*\rangle\simeq \frac{N_d\ln N_d}{\ln(z-1)}.
\end{equation}
Hence, the optimized MFPT grows slightly faster than linearly with $N_d$, the minimal size of the sub-tree to be explored to find the target. It is instructive to compare this time with the average time $ \langle T_d^{(\mathrm{syst})}\rangle$ it would take to find the target by using a systematic search strategy, consisting in visiting only once each site located at a distance $d$ from the origin, without going further than $d$. This systematic search is the best possible strategy (if the searcher is informed that the target is located at a distance $d$). The minimal total number of steps necessary to visit all the sites at a distance $d$ one by one is twice the number of links $L_d$ of the Cayley tree with $N_d$ nodes. This can be understood by noting that the walker needs to cross a link once on its way to the boundary and once on its way back toward the origin. Since on average, the target will be found after visiting half of the nodes at a distance $d$, $\langle T_d^{(\mathrm{syst})}\rangle=2L_d/2=L_d$. Noting that $L_d\simeq N_d$ for large Cayley trees, we obtain
\begin{equation}
  \langle T_d^{(\mathrm{syst})}\rangle\simeq N_d.   
\end{equation}
We hence come to the conclusion that the random walk with optimized resetting will take only $\ln N_d$ (or $d$) times longer than the best possible strategy
\begin{equation}
    \frac{\langle T_d^*\rangle}{\langle T_d^{(\mathrm{syst})}\rangle}\simeq 
    \frac{\ln N_d}{\ln(z-1)}\simeq d.
\end{equation}
On regular $D$-dimensional lattices, this multiplicative factor is much larger, of $O(N_d^{1/D})$ \cite{evans2014diffusion}. Hence, searches with optimized resetting are very efficient on Cayley trees, and possibly on other large networks where the number of nodes increases exponentially with the distance, which is the case of many complex networks.
\\[2mm]
\section{Random and complex networks} 
With the help of Eqs. (\ref{Pinfvectors})-(\ref{globalMFPT}), we further analyze different types of networks of relatively small size ($N=100$) for clarity in the visualizations. 

Figure \ref{Figure4}(a) displays the global time 
$\mathrm{T}(i,\gamma)$ as a function of $\gamma$ on a Barbell graph, {\it i.e.}, a network model with two well-defined communities composed of two fully connected networks (of $45$ nodes each) connected by a chain (of $10$ nodes) \cite{Ghosh2008Barbell}. Whereas network exploration depends remarkably little on the initial node for the simple random walk ($\gamma=0)$,
it becomes extremely sensitive to the position of $i$ as soon as resetting is switched on.
A moderate resetting probability can either increase or reduce $\mathrm{T}(i,\gamma)$ by orders of magnitude depending on the centrality $C_i$ of the starting node. Network exploration becomes very efficient and can be optimized at a non-zero resetting probability when one chooses a resetting node of high centrality, that lies in-between the two communities.
\\[2mm]
Figure \ref{Figure4}(b) shows qualitatively similar results for a Watts-Strogatz network \cite{WattsStrogatz1998} generated from a ring with nearest-neighbor and next-nearest-neighbour links and a rewiring probability of $p=0.01$. The shortcuts break the translational invariance of the ring geometry and the resetting nodes of higher
closeness centrality $C_i$, those close to a shortcut, tend to produce lower global MFPT, although the trend is less marked than in the previous example. 
\\[2mm]
The network in Fig. \ref{Figure4}(c) is the giant component of an Erd\"{o}s-R\'enyi (ER) random network  \cite{ErdosRenyi1959} with Poisson degree distribution and average degree $\langle k \rangle= 2.72$. Conversely, Fig. \ref{Figure4}(d) corresponds to a scale-free Barab\'asi-Albert (BA) network with power-law distributed node degrees, generated with the preferential attachment rule \cite{BarabasiAlbert1999}. As in the previous cases, the more peripheral resetting nodes (in red) usually cause a monotonous increase of the global MFPT with $\gamma$, whereas for the central nodes a minimum may exist. This situation is similar to the one described for diffusion with resetting in 1D bounded domains with reflective boundaries \cite{christou2015diffusion}. These examples also illustrate that the degree $k_i$ of the starting node plays a lesser role. The value of $C_i$ alone does not determine the shape of the MFPT, as network exploration is sensitive to the network architecture. For instance, when the network diameter is small, as it is the case for ER and BA networks, differences between the nodes tend to mitigate.
\\[2mm]
\section{Conclusions}
We have explored a stochastic process on networks that combines random walk steps to adjacent nodes and resetting to the initial node. Our formalism analyzes the dynamics in terms of the spectral representation of the transition matrix that defines the random walk strategy without resetting. We apply these results to characterize the dynamics on rings,  Cayley trees, and random networks, including scale-free and small-world networks. In Cayley trees, and possibly in many networks with few loops, the walk with an optimized resetting probability perform nearly as well as the best possible search strategy to find a target at a given distance. In a simple network model with communities, the efficiency of searches under reset can be increased or decreased by orders of magnitude, depending on the centrality of the resetting node. These results indicate that processes with resetting are promising strategies for exploring complex networks. The methods introduced are general and pave the way to further extensions of the study of resetting processes, which may be useful to investigate the structure of complex networks.
\\[2mm]
\section{Appendix}
\subsection{General properties} 
\label{AppendixA}
\subsubsection{Eigenvalues and eigenvectors of $\mathbf{\Pi}(r,\gamma)$}
We analyze the eigenvalues and eigenvectors of the matrix
\begin{equation}
\mathbf{\Pi}(r,\gamma)=(1-\gamma)\mathbf{W}+\gamma\mathbf{\Theta}(r),
\end{equation}
where $r=1,\ldots,N$ is the node to which resetting occurs with probability $0\leq \gamma<1$. The elements $\ell,m$ of the matrix $\mathbf{\Theta}(r)$ are $\Theta_{\ell m}(r)=\delta_{mr}$.
We express the results  in terms of the left and right eigenvectors $\{\left\langle\bar{\phi}_\ell\right|\}_{\ell= 1}^N$ , $\{\left|\phi_\ell\right\rangle \}_{\ell= 1}^N$ of the transition matrix $\mathbf{W}$ with eigenvalues $\{ \lambda_\ell\}_{\ell= 1}^N$. We have $\mathbf{W}\left|\phi_\ell\right\rangle=\lambda_\ell\left|\phi_\ell\right\rangle $ and $\left\langle\bar{\phi}_\ell\right|\mathbf{W}=\lambda_\ell \left\langle\bar{\phi}_\ell\right|$ for $\ell=1,\ldots,N$, where the set of eigenvalues is ordered in the form $\lambda_1=1$ and $1>\lambda_l\geq -1$ for $l=2,3,\ldots,N$. We define $|i\rangle$ as the vector whose components are 0 except the $i$-th one, which is 1. In the following we denote as  $\{\left| i \right\rangle \}_{i= 1}^N$ the canonical base of $\mathbb{R}^N$. 

With the right eigenvectors we define a matrix $\mathbf{Z}$ with elements $Z_{ij}=\left\langle i|\phi_j\right\rangle$. The matrix $\mathbf{Z}$ is invertible, and a new set of vectors $\left\langle \bar{\phi}_i\right|$ is obtained by $(\mathbf{Z}^{-1})_{ij}=\left\langle \bar{\phi}_i |j\right\rangle $, then
\begin{equation}\label{conditionOC}
\delta_{ij}=(\mathbf{Z}^{-1}\mathbf{Z})_{ij}=\sum_{\ell=1}^N \left\langle\bar{\phi}_i|\ell\right\rangle \left\langle \ell|\phi_j\right\rangle=\langle\bar{\phi}_i|\phi_j\rangle \,
\end{equation}
and
\begin{equation}\label{conditionC}
\mathbb{1}=\mathbf{Z}\mathbf{Z}^{-1}=\sum_{\ell=1}^N \left|\phi_\ell\right\rangle \left\langle \bar{\phi}_\ell \right| \, ,
\end{equation}
where $\mathbb{1}$ is the $N\times N$ identity matrix. In addition, by normalization of the probability, the matrix $\mathbf{W}$ is such that $\sum_{\ell=1}^N w_{i \to \ell}=1$, which implies that $|\phi_1 \rangle\propto \left(\begin{array}{c} 1\\ 1 \\ \ldots \\ 1\end{array}\right)$. 
\\[3mm]
By using $\left\langle\bar{\phi}_\ell|\phi_1\right\rangle=\sum_{i=1}^N \left\langle\bar{\phi}_\ell| i\right\rangle \left\langle i|\phi_1\right\rangle=\delta_{\ell 1}$ and considering the vector $\left\langle i|\phi_1\right\rangle=\left\langle r|\phi_1\right\rangle=\mathrm{constant}$ for $r=1,\ldots,N$; we obtain
\begin{equation}\label{conditionV1}
\sum_{i=1}^N \left\langle\bar{\phi}_\ell| i\right\rangle=\frac{\delta_{\ell 1}}{\left\langle r|\phi_1\right\rangle}.
\end{equation}
Hence, from relations in Eqs. (\ref{conditionOC})-(\ref{conditionV1}), we have
\begin{align}
\nonumber
\mathbf{\Theta}&(r)=\sum_{\ell= 1}^N\sum_{m=1}^N \left|\phi_\ell\right\rangle \left\langle\bar{\phi}_\ell\right| \mathbf{\Theta}(r)\left|\phi_m\right\rangle \left\langle\bar{\phi}_m\right|\\\nonumber
&=\sum_{\ell= 1}^N\sum_{m=1}^N \sum_{u=1}^N\sum_{v=1}^N \left|\phi_\ell\right\rangle \left\langle \bar{\phi}_\ell\right|u\rangle\left\langle u\right| \mathbf{\Theta}(r)|v\rangle\left\langle v|\phi_m\right\rangle \left\langle\bar{\phi}_m\right|\\  \nonumber
&=\sum_{\ell= 1}^N\sum_{m=1}^N \sum_{u=1}^N\sum_{v=1}^N \left|\phi_\ell\right\rangle \left\langle \bar{\phi}_\ell\right|u\rangle\delta_{vr} \left\langle v |\phi_m\right\rangle \left\langle\bar{\phi}_m\right|\\\nonumber
&=\sum_{\ell= 1}^N\sum_{m=1}^N \sum_{u=1}^N \left|\phi_\ell\right\rangle \left\langle \bar{\phi}_\ell\right|u\rangle\left\langle r |\phi_m\right\rangle \left\langle\bar{\phi}_m\right|\\\nonumber
&=\sum_{\ell= 1}^N\sum_{m=1}^N  \left|\phi_\ell\right\rangle \left[\sum_{u=1}^N \left\langle \bar{\phi}_\ell\right|u\rangle\right]\left\langle r |\phi_m\right\rangle \left\langle\bar{\phi}_m\right|\\
&=
\sum_{\ell= 1}^N\sum_{m=1}^N  \left|\phi_\ell\right\rangle \frac{\delta_{\ell 1}}{\left\langle r|\phi_1\right\rangle}\left\langle r |\phi_m\right\rangle \left\langle\bar{\phi}_m\right|.
\end{align}
Therefore,
\begin{equation}\label{Thetaphirep}
\mathbf{\Theta}(r)=\sum_{m=1}^N   \frac{\left\langle r |\phi_m\right\rangle }{\left\langle r|\phi_1\right\rangle}\left|\phi_1\right\rangle\left\langle\bar{\phi}_m\right|.
\end{equation}
In the following, we explore the right and left eigenvectors of $\mathbf{\Pi}(r,\gamma)$, denoted as $\left|\psi_\ell(r,\gamma)\right\rangle$ and $\left\langle\bar{\psi}_\ell(r,\gamma)\right|$, and satisfying the relations
\begin{align*}
 \mathbf{\Pi}(r,\gamma)\left|\psi_\ell(r,\gamma)\right\rangle&=\zeta_\ell(r,\gamma)\left|\psi_\ell(r,\gamma)\right\rangle,\\
\left\langle\bar{\psi}_\ell(r,\gamma)\right|\mathbf{\Pi}(r,\gamma)&=\zeta_\ell(r,\gamma)\left\langle\bar{\psi}_\ell(r,\gamma)\right|
\end{align*}
for $\ell=1,2,\ldots,N$, where the eigenvalues of $\mathbf{\Pi}(r,\gamma)$ are $\zeta_\ell(r,\gamma)$.
From the result in Eq. (\ref{Thetaphirep}), we see that $\mathbf{\Theta}(r)\left| \phi_1\right\rangle=\left| \phi_1\right\rangle$. Therefore, one sees that $\left|\psi_1(r,\gamma)\right\rangle=\left| \phi_1\right\rangle$, since
\begin{align}\nonumber
\mathbf{\Pi}(r,\gamma)&\left|\phi_1\right\rangle=\left[(1-\gamma)\mathbf{W}+\gamma \mathbf{\Theta}(r)\right]\left|\phi_1\right\rangle\\\nonumber
&=(1-\gamma)\left|\phi_1\right\rangle+\gamma \mathbf{\Theta}(r)\left|\phi_1\right\rangle\\ \nonumber
&=(1-\gamma)\left|\phi_1\right\rangle+\gamma \sum_{m=1}^N   \frac{\left\langle r |\phi_m\right\rangle }{\left\langle r|\phi_1\right\rangle}\left|\phi_1\right\rangle\left\langle\bar{\phi}_m|\phi_1\right\rangle\\\nonumber
&=(1-\gamma)\left|\phi_1\right\rangle+\gamma \sum_{m=1}^N \frac{\left\langle r |\phi_m\right\rangle }{\left\langle r|\phi_1\right\rangle}\left|\phi_1\right\rangle\delta_{m1}\\
&=(1-\gamma)\left|\phi_1\right\rangle+\gamma\left|\phi_1\right\rangle=\left|\phi_1\right\rangle\\\nonumber
&=\left|\psi_1(r,\gamma)\right\rangle=\zeta_1(r,\gamma)\left|\psi_1(r,\gamma)\right\rangle
\end{align}
where $\zeta_1(r,\gamma)=1$.
\\[3mm]
In a similar way, we see that $\left\langle\bar{\phi}_\ell\right| \mathbf{\Theta}(r)=0$ for $\ell=2,3,\ldots,N$. As a consequence,  we deduce that $\left\langle\bar{\psi}_\ell(r,\gamma)\right|=\left\langle\bar{\phi}_\ell\right|$ for
$\ell=2,3,\ldots,N$, since
\begin{align}\nonumber
\left\langle\bar{\phi}_\ell\right|&\mathbf{\Pi}(r,\gamma)= \left\langle\bar{\phi}_\ell\right|  \left[(1-\gamma)\mathbf{W}+\gamma \mathbf{\Theta}(r)\right]\\\nonumber
&=(1-\gamma)\lambda_\ell \left\langle\bar{\phi}_\ell\right|  +\gamma \left\langle\bar{\phi}_\ell\right| \mathbf{\Theta}(r)\\ \nonumber
&=(1-\gamma)\lambda_\ell \left\langle\bar{\phi}_\ell\right|  +\gamma  \sum_{m=1}^N   \frac{\left\langle r |\phi_m\right\rangle }{\left\langle r|\phi_1\right\rangle}\left\langle\bar{\phi}_\ell|\phi_1\right\rangle\left\langle\bar{\phi}_m\right|\\\nonumber
&=(1-\gamma)\lambda_\ell \left\langle\bar{\phi}_\ell\right|  +\gamma  \sum_{m=1}^N   \frac{\left\langle r |\phi_m\right\rangle }{\left\langle r|\phi_1\right\rangle}\delta_{\ell 1}\left\langle\bar{\phi}_m\right|\\
&=(1-\gamma)\lambda_\ell \left\langle\bar{\phi}_\ell\right|=\zeta_\ell(r,\gamma)\left\langle\bar{\psi}_\ell(r,\gamma)\right|.
\end{align}
This result shows that $\left\langle\bar{\psi}_\ell(r,\gamma)\right|=\left\langle\bar{\phi}_\ell\right|$ and $\zeta_\ell(r,\gamma)=(1-\gamma)\lambda_\ell$ for $\ell=2,3,\ldots,N$.
\\[3mm]
Now, we deduce the rest of the eigenvectors. For $\left\langle\bar{\psi}_1(r,\gamma)\right|$, we use the ansatz
\begin{equation}
\left\langle\bar{\psi}_1(r,\gamma)\right|=\left\langle\bar{\phi}_1\right|+\sum_{m=2}^N a_m\left\langle\bar{\phi}_m\right|.
\end{equation}
This choice is motivated by the structure of the matrix $\mathbf{\Theta}(r)$ in Eq. (\ref{Thetaphirep}). Here, the goal is to deduce the values $\{a_m\}_{m=2}^N$. We know that $\left\langle\bar{\psi}_1(r,\gamma)\right|\mathbf{\Pi}(r,\gamma)=\left\langle\bar{\psi}_1(r,\gamma)\right|$. Therefore
\begin{align}\nonumber
&\left\langle\bar{\psi}_1(r,\gamma)\right|\mathbf{\Pi}(r,\gamma)=\left(\left\langle\bar{\phi}_1\right|+\sum_{m=2}^N a_m\left\langle\bar{\phi}_m\right|\right)\\\nonumber
&\qquad\qquad\times\left[(1-\gamma)\mathbf{W}+\gamma\mathbf{\Theta}(r)\right]\\ \nonumber
&=(1-\gamma)\left\langle\bar{\phi}_1\right|+\gamma\sum_{m=1}^N \frac{\left\langle r |\phi_m\right\rangle }{\left\langle r|\phi_1\right\rangle}\left\langle\bar{\phi}_m\right|\\\nonumber
&\qquad\qquad+(1-\gamma)\sum_{m=2}^N a_m\,\lambda_m \left\langle\bar{\phi}_m\right|\\
&=
\left\langle\bar{\phi}_1\right|+\sum_{m=2}^N \left[\gamma\frac{\left\langle r |\phi_m\right\rangle }{\left\langle r|\phi_1\right\rangle}+(1-\gamma)a_m\,\lambda_m\right] \left\langle\bar{\phi}_m\right|.
\end{align}
This requires $a_m=\gamma\frac{\left\langle r |\phi_m\right\rangle }{\left\langle r|\phi_1\right\rangle}+(1-\gamma)a_m\,\lambda_m$. Therefore $
a_m=\frac{\gamma}{1-(1-\gamma)\lambda_m}\frac{\left\langle r |\phi_m\right\rangle }{\left\langle r|\phi_1\right\rangle}$. Hence, we have
\begin{equation}
\left\langle\bar{\psi}_1(r,\gamma)\right|=\left\langle\bar{\phi}_1\right|+\sum_{m=2}^N \frac{\gamma}{1-(1-\gamma)\lambda_m}\frac{\left\langle r |\phi_m\right\rangle }{\left\langle r|\phi_1\right\rangle}\left\langle\bar{\phi}_m\right|.
\end{equation}
Finally, we consider the eigenvectors $\left|\psi_\ell(r,\gamma)\right\rangle$ for $\ell=2,3,\ldots,N$. From Eq. (\ref{Thetaphirep}) we know that $\mathbf{\Theta}(r)\left| \phi_\ell\right\rangle=\frac{\left\langle r |\phi_\ell\right\rangle }{\left\langle r|\phi_1\right\rangle}\left| \phi_1\right\rangle$, which motivates the ansatz
\begin{equation}
\left|\psi_\ell(r,\gamma)\right\rangle=\left| \phi_\ell\right\rangle+b_\ell \left| \phi_1\right\rangle\qquad\mathrm{for}\quad \ell=2,3,\ldots,N.
\end{equation}
Since $\mathbf{\Pi}(r,\gamma)\left|\psi_\ell(r,\gamma)\right\rangle=(1-\gamma)\lambda_l \left|\psi_\ell(r,\gamma)\right\rangle$, we have
\begin{align}\nonumber
&\mathbf{\Pi}(r,\gamma)\left|\psi_\ell(r,\gamma)\right\rangle=\left[(1-\gamma)\mathbf{W}+\gamma \mathbf{\Theta}(r)\right]\left(\left| \phi_\ell\right\rangle+b_\ell \left| \phi_1\right\rangle\right)\\\nonumber
&=(1-\gamma)\lambda_\ell\left|\phi_\ell\right\rangle+\gamma \frac{\left\langle r |\phi_\ell\right\rangle }{\left\langle r|\phi_1\right\rangle}\left| \phi_1\right\rangle+(1-\gamma) b_\ell\left| \phi_1\right\rangle+\gamma b_\ell\left| \phi_1\right\rangle\\ 
&=(1-\gamma)\lambda_\ell\left[
\left|\phi_\ell\right\rangle+\frac{1}{(1-\gamma)\lambda_\ell}\left(b_\ell+\gamma \frac{\left\langle r |\phi_\ell\right\rangle }{\left\langle r|\phi_1\right\rangle}  \right)\left| \phi_1\right\rangle
\right].
\end{align}
By identification, $b_\ell=\frac{1}{(1-\gamma)\lambda_\ell}\left(b_\ell+\gamma \frac{\left\langle r |\phi_\ell\right\rangle }{\left\langle r|\phi_1\right\rangle}  \right)$, therefore $b_\ell=-\frac{\gamma}{1-(1-\gamma)\lambda_\ell}\frac{\left\langle r |\phi_\ell\right\rangle }{\left\langle r|\phi_1\right\rangle}$. Then, for $\ell=2,3,\ldots,N$
\begin{equation}
\left|\psi_\ell(r,\gamma)\right\rangle=\left| \phi_\ell\right\rangle-\frac{\gamma}{1-(1-\gamma)\lambda_\ell}\frac{\left\langle r |\phi_\ell\right\rangle }{\left\langle r|\phi_1\right\rangle} \left| \phi_1\right\rangle.
\end{equation}
In summary, for the transition matrix $\mathbf{\Pi}(r,\gamma)$, we obtained the set of right eigenvectors
\begin{equation}\label{ReigVPhi}
\left|\psi_\ell(r,\gamma)\right\rangle=
\begin{cases}
\displaystyle
\left| \phi_1\right\rangle&\ell= 1,\\
\left|\phi_\ell\right\rangle-\frac{\gamma\frac{\left\langle r|\phi_\ell\right\rangle }{\left\langle r|\phi_1\right\rangle}}{1-(1-\gamma)\lambda_\ell}\left|\phi_1\right\rangle  &\ell= 2,\ldots,N, \\
\end{cases}
\end{equation}
and the left eigenvectors
\begin{equation}\label{LeigVPhi}
\left\langle\bar{\psi}_\ell(r,\gamma)\right|=
\begin{cases}
\left\langle\bar{\phi}_1\right|
+\sum\limits_{m=2}^N\frac{\gamma\frac{\left\langle r|\phi_m\right\rangle}{\left\langle r|\phi_1\right\rangle}}{1-(1-\gamma)\lambda_m}\left\langle\bar{\phi}_m\right|,\ell=1,\\
\left\langle\bar{\phi}_\ell\right|\hspace{2.5cm}   \ell=2,\ldots,N, \\
\end{cases}
\end{equation}
with eigenvalues
\begin{equation}\label{eigvalsPi}
\zeta_\ell(r,\gamma)=
\begin{cases}
\displaystyle
1 \quad &\mathrm{for}\quad \ell= 1,\\
(1-\gamma)\lambda_\ell \quad &\mathrm{for}\quad \ell= 2,3,\ldots, N.
\end{cases}
\end{equation}
\subsubsection{Orthonormalization and completeness relation}
Now we check the orthonormalization property $\left\langle\bar{\psi}_\ell(r,\gamma)|\psi_m(r,\gamma)\right\rangle=\delta_{\ell m}$ and the completeness relation $\sum_{\ell= 1}^N \left|\psi_\ell\right\rangle \left\langle\bar{\psi}_\ell\right|=\mathbb{1}$ satisfied by the eigenvectors of  $\mathbf{\Pi}(r,\gamma)$ in Eqs. (\ref{ReigVPhi}) and (\ref{LeigVPhi}).
\\[4mm]
We start with the completeness relation $\sum_{\ell= 1}^N \left|\psi_\ell(r,\gamma)\right\rangle \left\langle\bar{\psi}_\ell(r,\gamma)\right|=\mathbb{1}$, we have
\begin{align}\nonumber
&\sum_{\ell= 1}^N \left|\psi_\ell(r,\gamma)\right\rangle \left\langle\bar{\psi}_\ell(r,\gamma)\right|=\left|\psi_1(r,\gamma)\right\rangle \left\langle\bar{\psi}_1(r,\gamma)\right|\\ \nonumber
&\qquad+\sum_{\ell= 2}^N \left|\psi_\ell(r,\gamma)\right\rangle \left\langle\bar{\psi}_\ell(r,\gamma)\right|\\ \nonumber
&=\left|\phi_1\right\rangle \left\langle\bar{\phi}_1\right|+\sum_{m=2}^N\frac{\gamma}{1-(1-\gamma)\lambda_m}\frac{\left\langle r|\phi_m\right\rangle}{\left\langle r|\phi_1\right\rangle}\left|\phi_1\right\rangle \left\langle\bar{\phi}_m\right| \\ \nonumber
&\qquad+ \sum_{\ell= 2}^N \left|\phi_\ell\right\rangle \left\langle\bar{\phi}_\ell\right|-
\sum_{\ell= 2}^N \frac{\gamma}{1-(1-\gamma)\lambda_\ell}\frac{\left\langle r|\phi_\ell\right\rangle}{\left\langle r|\phi_1\right\rangle}\left|\phi_1\right\rangle \left\langle\bar{\phi}_\ell\right|   \\
&=\left|\phi_1\right\rangle \left\langle\bar{\phi}_1\right|+\sum_{\ell= 2}^N \left|\phi_\ell\right\rangle \left\langle\bar{\phi}_\ell\right|=\sum_{\ell= 1}^N \left|\phi_\ell\right\rangle \left\langle\bar{\phi}_\ell\right|=\mathbb{1}.
\end{align}
Now, let us check that  $\left\langle\bar{\psi}_\ell(r,\gamma)|\psi_m(r,\gamma)\right\rangle=\delta_{\ell m}$. We have the following cases:
\\[4mm]
$\bullet$ \hspace{5mm} Calculation of $\left\langle\bar{\psi}_1(r,\gamma)|\psi_1(r,\gamma)\right\rangle$:
\begin{align}\\ \nonumber
&\left\langle\bar{\psi}_1(r,\gamma)|\psi_1(r,\gamma)\right\rangle=\left\langle\bar{\phi}_1|\phi_1\right\rangle\\ \nonumber
&\qquad+\sum_{m=2}^N\frac{\gamma}{1-(1-\gamma)\lambda_m}\frac{\left\langle r|\phi_m\right\rangle}{\left\langle r|\phi_1\right\rangle}\left\langle\bar{\phi}_m|\phi_1\right\rangle\\ 
&=1+\sum_{m=2}^N\frac{\gamma}{1-(1-\gamma)\lambda_m}\frac{\left\langle r|\phi_m\right\rangle}{\left\langle r|\phi_1\right\rangle}\delta_{m1}=1.
\end{align}
$\bullet$ \hspace{5mm} Calculation of $\left\langle\bar{\psi}_1(r,\gamma)|\psi_\ell(r,\gamma)\right\rangle$ for $\ell=2,3,\ldots,N$:
\begin{align}\nonumber
&\left\langle\bar{\psi}_1(r,\gamma)|\psi_\ell(r,\gamma)\right\rangle\\ \nonumber
&=\left[\left\langle\bar{\phi}_1\right|
+\gamma\sum_{m=2}^N\frac{1}{1-(1-\gamma)\lambda_m}\frac{\left\langle r|\phi_m\right\rangle}{\left\langle r|\phi_1\right\rangle}\left\langle\bar{\phi}_m\right| \right]\\ \nonumber
&\qquad\times\left[\left|\phi_\ell\right\rangle-\frac{\gamma}{1-(1-\gamma)\lambda_\ell}\frac{\left\langle r|\phi_\ell\right\rangle }{\left\langle r|\phi_1\right\rangle}\left|\phi_1\right\rangle\right]\\
\nonumber
&=\left\langle\bar{\phi}_1|\phi_\ell\right\rangle+\sum_{m=2}^N\frac{\gamma}{1-(1-\gamma)\lambda_m}\frac{\left\langle r|\phi_m\right\rangle}{\left\langle r|\phi_1\right\rangle}\\ \nonumber
&\qquad\times\left[\left\langle\bar{\phi}_m|\phi_\ell\right\rangle-\frac{\gamma}{1-(1-\gamma)\lambda_\ell}\frac{\left\langle r|\phi_\ell\right\rangle}{\left\langle r|\phi_1\right\rangle}\left\langle\bar{\phi}_m|\phi_1\right\rangle\right]\\ \nonumber
&\qquad-\frac{\gamma}{1-(1-\gamma)\lambda_\ell}\frac{\left\langle r|\phi_\ell\right\rangle }{\left\langle r|\phi_1\right\rangle}.
\end{align}
Therefore
\begin{align}\nonumber
&\left\langle\bar{\psi}_1(r,\gamma)|\psi_\ell(r,\gamma)\right\rangle=\delta_{1\ell}\\\nonumber
&+\sum_{m=2}^N\frac{\gamma}{1-(1-\gamma)\lambda_m}\frac{\left\langle r|\phi_m\right\rangle}{\left\langle r|\phi_1\right\rangle}\\ \nonumber
&\qquad\times\left[\delta_{\ell m}-\frac{\gamma}{1-(1-\gamma)\lambda_\ell}\frac{\left\langle r|\phi_\ell\right\rangle}{\left\langle r|\phi_1\right\rangle}\delta_{m1}\right]\\
&\qquad-\frac{\gamma}{1-(1-\gamma)\lambda_\ell}\frac{\left\langle r|\phi_\ell\right\rangle }{\left\langle r|\phi_1\right\rangle}.
\end{align}
However, since  $\ell=2,3,\ldots,N$, we have
\begin{multline}
\left\langle\bar{\psi}_1(r,\gamma)|\psi_\ell(r,\gamma)\right\rangle=\frac{\gamma}{1-(1-\gamma)\lambda_\ell}\frac{\left\langle r|\phi_\ell\right\rangle}{\left\langle r|\phi_1\right\rangle}\\-\frac{\gamma}{1-(1-\gamma)\lambda_\ell}\frac{\left\langle r|\phi_\ell\right\rangle }{\left\langle r|\phi_1\right\rangle}=0\quad \mathrm{for}\,\, \ell= 2,3,\ldots,N.
\end{multline}

$\bullet$ \hspace{5mm} We have $\left\langle\bar{\psi}_\ell(r,\gamma)|\psi_1(r,\gamma)\right\rangle=0$ for $\ell=2,3,\ldots,N$, since:
\begin{equation}
\left\langle\bar{\psi}_\ell(r,\gamma)|\psi_1(r,\gamma)\right\rangle=\left\langle\bar{\phi}_\ell(r,\gamma)|\phi_1(r,\gamma)\right\rangle=\delta_{\ell 1}=0.
\end{equation}
$\bullet$ \hspace{5mm} We have $\left\langle\bar{\psi}_\ell(r,\gamma)|\psi_m(r,\gamma)\right\rangle=\delta_{\ell m}$ for $\ell,m=2,3,\ldots,N$, since:
\begin{align}\nonumber
&\left\langle\bar{\psi}_\ell(r,\gamma)|\psi_m(r,\gamma)\right\rangle\\\nonumber
&=\left\langle\bar{\phi}_\ell\right|\left[\left|\phi_m\right\rangle-\frac{\gamma}{1-(1-\gamma)\lambda_m}\frac{\left\langle r|\phi_m\right\rangle }{\left\langle r|\phi_1\right\rangle}\left|\phi_1\right\rangle\right]\\
&=
\delta_{\ell m}-\frac{\gamma}{1-(1-\gamma)\lambda_m}\frac{\left\langle r|\phi_m\right\rangle }{\left\langle r|\phi_1\right\rangle}\delta_{\ell 1} =\delta_{\ell m}
\end{align}
for $\ell,m=2,3,\ldots,N$.
The results presented in this section prove that relations in Eqs. (\ref{ReigVPhi})-(\ref{eigvalsPi}) define the eigenvalues and eigenvectors of the transition matrix $\mathbf{\Pi}(r,\gamma)$ that describes the dynamics with resetting to the node $r$. The sets composed of the left and right eigenvectors form an orthonormalized base, a result that allows us to deduce analytical expressions for different quantities of interest for the dynamics of a random walker with resetting.

\subsubsection{Stationary distribution and mean first passage time}\label{sec:mfptApp}
In this part we present the general expressions for the occupation probabilities and mean first passage times. We center our discussion on the analysis of a Markovian process defined by the transition matrix $\mathbf{\Pi}(r,\gamma)$, and then specify the results for random walks with resetting on networks. The occupation probability $P_{ij}(t;r,\gamma)$ can be expressed as \cite{Hughes,NohRieger2004}
\begin{equation}\label{EquF}
P_{ij}(t;r,\gamma)  = \delta_{t0} \delta_{ij} + \sum_{t'=0}^t   P_{jj}(t-t';r,\gamma)  F_{ij}(t';r,\gamma) \ ,
\end{equation}
where $F_{ij}(t;r,\gamma)$ is the probability of finding the process at $j$ for the first time after $t$ steps, starting from $i$. Using the discrete Laplace transform $\tilde{f}(s) \equiv\sum_{t=0}^\infty e^{-st} f(t)$ in Eq. (\ref{EquF}) we have \cite{NohRieger2004} 
\begin{equation}\label{LaplTransF}
\widetilde{F}_{ij} (s;r,\gamma) = \frac{\widetilde{P}_{ij}(s;r,\gamma) - \delta_{ij}}{\widetilde{P}_{jj} (s;r,\gamma)} .
\end{equation}
The mean first passage time (MFPT) $\langle T_{ij}(r,\gamma)\rangle$, defined as the mean number of steps taken to reach $j$ for the first time, starting from $i$ \cite{Hughes}, can be obtained through the series expansion of  $\widetilde{F}_{ij} (s;r,\gamma)$
\begin{equation}
\widetilde{F}_{ij} (s;r,\gamma)=1-s\langle T_{ij}(r,\gamma)\rangle+\frac{s^2}{2}\langle T^2_{ij}(r,\gamma)\rangle+\ldots, 
\end{equation}
where $\langle T^2_{ij}(r,\gamma)\rangle$ is the second moment of $F_{ij}(t;r,\gamma)$, or the ensemble average (over trajectories) of the squares of the first passage time from $i$ to $j$. In addition, the stationary distribution $P_j^\infty(r,\gamma)$ of the process is defined as 
\begin{equation}\label{Pinfdef}
P_j^\infty(r,\gamma)\equiv\lim_{T\to \infty}\frac{1}{T}\sum_{t=0}^T P_{ij}(t;r,\gamma),
\end{equation}
which is assumed to be independent of the initial condition and gives the probability to occupy $j$ when $t\to \infty$. Now, given $P_j^\infty(r,\gamma)$, let us define the moments 
$R^{(n)}_{ij}(r,\gamma)\equiv \sum_{t=0}^{\infty} t^n ~ \{P_{ij}(t;r,\gamma)-P_j^\infty(r,\gamma)\}$. The expansion in series of $\widetilde{P}_{ij}(s;r,\gamma)$ can be recast as
\begin{equation}
\widetilde{P}_{ij}(s;r,\gamma) = \frac{P_j^\infty(r,\gamma)}{(1-e^{-s})}
+ \sum_{n=0}^\infty (-1)^n R^{(n)}_{ij}(r,\gamma) \frac{s^n}{n!} \ .
\end{equation}
Substituting this result into Eq. (\ref{LaplTransF}) and performing a series expansion of $\widetilde{F}_{ij} (s;r,\gamma)$, we obtain, by identification
\begin{equation}\label{Tij_R}
\langle T_{ij}(r,\gamma)\rangle=\frac{R_{jj}^{(0)}(r,\gamma)-R_{ij}^{(0)}(r,\gamma)+\delta _{ij}}{P_j^{\infty}(r,\gamma)}.
\end{equation}
To further calculate  $P_j^\infty(r,\gamma)$ and $\langle T_{ij}(r,\gamma)\rangle$, we need to obtain $P_{ij}(t;r,\gamma)$. We start with the matrix form of the master equation $\vec{P}(t;r,\gamma)=\vec{P}(0)\mathbf{\Pi}(r,\gamma)^t$ where $\vec{P}(t;r,\gamma)$ is the probability vector at time $t$. Using Dirac's notation
\begin{equation}
P_{ij}(t;r,\gamma)=\left\langle i\right|\mathbf{\Pi}(r,\gamma)^t \left|j\right\rangle,
\end{equation}
where $\{\left|m\right\rangle \}_{m=1}^N$ represents the canonical base of $\mathbb{R}^N$.  In terms of the eigenvectors and eigenvalues of $\mathbf{\Pi}(r,\gamma)$, we have the spectral representation
\begin{equation}\label{ProbVector}
\mathbf{\Pi}(r,\gamma)=\sum_{\ell=1}^N \zeta_\ell(r,\gamma) \left|\psi_\ell(r,\gamma)\right\rangle \left\langle\bar{\psi}_\ell(r,\gamma)\right|. 
\end{equation}
The spectral form of the transition matrix in Eq. (\ref{ProbVector}) allows us to obtain $P_{ij}(t)$
\begin{equation}
P_{ij}(t;r,\gamma)=\sum_{\ell=1}^N [\zeta_\ell(r,\gamma)]^t \left\langle i|\psi_\ell(r,\gamma)\right\rangle \left\langle\bar{\psi}_\ell(r,\gamma)|j\right\rangle. 
\end{equation}
Therefore,  the stationary distribution $P_{j}^\infty(r,\gamma)$ in Eq. (\ref{Pinfdef}) is
\begin{align}\nonumber
&P_{j}^\infty(r,\gamma)\\\nonumber
&= \lim_{T\to \infty} \frac{1}{T}\sum_{t=0}^T\sum_{\ell=1}^N(\zeta_\ell(r,\gamma))^t \left\langle i|\psi_\ell(r,\gamma)\right\rangle \left\langle\bar{\psi}_\ell(r,\gamma)|j\right\rangle\\ \nonumber
&= \sum_{\ell=1}^N \left[ \lim_{T\to\infty}\frac{1}{T}\sum_{t=0}^T (\zeta_\ell(r,\gamma))^t\right]\left\langle i|\psi_\ell(r,\gamma)\right\rangle \left\langle\bar{\psi}_\ell(r,\gamma)|j\right\rangle \\\nonumber
&=\zeta_1(r,\gamma)\left\langle i|\psi_1(r,\gamma)\right\rangle\left\langle \bar{\psi}_1(r,\gamma)|j\right\rangle
\\
&=\left\langle i|\psi_1(r,\gamma)\right\rangle\left\langle \bar{\psi}_1(r,\gamma)|j\right\rangle.
\label{StatEigvec1}
\end{align}
The property $\left\langle i|\psi_1(r,\gamma)\right\rangle=\mathrm{constant}$ makes the stationary distribution $P_{j}^\infty(r,\gamma)$ independent of the initial position. In a similar way, by using the definition of $R_{ij}^{(0)}(r,\gamma)$, we have
\begin{align}\nonumber
&R_{ij}^{(0)}(r,\gamma)=\sum_{t=0}^\infty (P_{ij}(t;r,\gamma)-P_j^{\infty}(r,\gamma))\\\nonumber
&=\sum_{t=0}^\infty \sum_{\ell=2}^N [\zeta_\ell(r,\gamma)]^t \left\langle i|\psi_\ell(r,\gamma)\right\rangle \left\langle\bar{\psi}_\ell(r,\gamma)|j\right\rangle\\
&=\sum_{\ell=2}^N\frac{1}{1-\zeta_\ell(r,\gamma)}\left\langle i|\psi_\ell(r,\gamma)\right\rangle \left\langle\bar{\psi}_\ell(r,\gamma)|j\right\rangle.
\label{Rij0}
\end{align}
Substitution of Eq. (\ref{Rij0}) into (\ref{Tij_R}) yields
\begin{multline}\label{TijSpect}
\left\langle T_{ij}(r,\gamma)\right\rangle
=\frac{1}{P_j^\infty(r,\gamma)}\bigl[\delta_{ij}\\+\sum_{\ell=2}^N\frac{1}{1-\zeta_\ell(r,\gamma)}\bigl(\left\langle j|\psi_\ell(r,\gamma)\right\rangle \left\langle\bar{\psi}_\ell(r,\gamma)|j\right\rangle\\-\left\langle i|\psi_\ell(r,\gamma)\right\rangle \left\langle\bar{\psi}_\ell(r,\gamma)|j\right\rangle\bigr)\bigr].
\end{multline}
We now use our previous findings, in Eqs. (\ref{ReigVPhi})-(\ref{eigvalsPi}), that established a connection between the eigenvalues and eigenvectors of the matrix $\mathbf{\Pi}(r,\gamma)$ and the matrix $\mathbf{W}$ for a random walker without resetting. We obtain for the stationary distribution in Eq. (\ref{StatEigvec1})
\begin{align}\nonumber
&P_{j}^\infty(r,\gamma)=\left\langle i|\psi_1(r,\gamma)\right\rangle\left\langle \bar{\psi}_1(r,\gamma)|j\right\rangle\\\nonumber
&=\left\langle i|\phi_1\right\rangle\left[\left\langle\bar{\phi}_1|j\right\rangle
+\gamma\sum_{m=2}^N\frac{\frac{\left\langle r|\phi_m\right\rangle}{\left\langle r|\phi_1\right\rangle}}{1-(1-\gamma)\lambda_m}\left\langle\bar{\phi}_m|j\right\rangle
\right]\\
&=\left\langle i|\phi_1\right\rangle\left\langle\bar{\phi}_1|j\right\rangle
+\gamma\sum_{m=2}^N\frac{\left\langle r|\phi_m\right\rangle\left\langle\bar{\phi}_m|j\right\rangle}{1-(1-\gamma)\lambda_m}. \label{PinfvectorsSM}
\end{align}
Here, $\left\langle i|\phi_1\right\rangle\left\langle\bar{\phi}_1|j\right\rangle$ is the stationary distribution of the random walker without resetting. In the particular case of a standard random walker with transition probabilities $w_{i\to j}=\frac{A_{ij}}{k_i}$,  $\left\langle i|\phi_1\right\rangle\left\langle\bar{\phi}_1|j\right\rangle=\frac{k_j}{\sum_{m=1}^N k_m}$ \cite{NohRieger2004}.
\\[2mm]
For the mean first passage time, we have, for $\ell=2,\ldots,N$,
\begin{align}\nonumber
&\left\langle i|\psi_\ell(r,\gamma)\right\rangle \left\langle\bar{\psi}_\ell(r,\gamma)|j\right\rangle\\\nonumber
&=\left(\left\langle i|\phi_\ell\right\rangle-\frac{\gamma}{1-(1-\gamma)\lambda_\ell}\left\langle r|\phi_\ell\right\rangle\right)
\left\langle\bar{\phi}_\ell|j\right\rangle\\
&=\left\langle i|\phi_\ell\right\rangle
\left\langle\bar{\phi}_\ell|j\right\rangle-\frac{\gamma}{1-(1-\gamma)\lambda_\ell}\left\langle r|\phi_\ell\right\rangle
\left\langle\bar{\phi}_\ell|j\right\rangle.
\end{align}
Therefore
\begin{multline}\label{psirelation}
\left\langle j|\psi_\ell(r,\gamma)\right\rangle \left\langle\bar{\psi}_\ell(r,\gamma)|j\right\rangle-\left\langle i|\psi_\ell(r,\gamma)\right\rangle \left\langle\bar{\psi}_\ell(r,\gamma)|j\right\rangle\\
=\left\langle j|\phi_\ell\right\rangle\left\langle\bar{\phi}_\ell|j\right\rangle-\left\langle i|\phi_\ell\right\rangle\left\langle\bar{\phi}_\ell|j\right\rangle \qquad \ell=2,\ldots,N.
\end{multline}
This expression is independent of the node $r$ and of the probability $\gamma$. Substituting Eq. (\ref{psirelation}) into (\ref{TijSpect}), we obtain $\left\langle T_{ij}(r,\gamma)\right\rangle$:
\begin{multline}\label{MFPT_resetSM}
\left\langle T_{ij}(r,\gamma)\right\rangle=\frac{\delta_{ij}}{P_j^\infty(r,\gamma)}\\+
\frac{1}{P_j^\infty(r,\gamma)}\sum_{\ell=2}^N\frac{
\left\langle j|\phi_\ell\right\rangle \left\langle\bar{\phi}_\ell|j\right\rangle-\left\langle i|\phi_\ell\right\rangle \left\langle\bar{\phi}_\ell|j\right\rangle
}{1-(1-\gamma)\lambda_\ell}.
\end{multline}
\subsection{Continuous-time random walks}
\label{AppendixB}

We now discuss the relation between the discrete-time random walker with resetting defined in terms of the transition matrix $\mathbf{\Pi}(r,\gamma)$ and the continuous-time version of this dynamics. Considering that each step in the discrete case is performed with regular time increments $\Delta t$, the master equation becomes
\begin{equation}
p_{ij}(t+\Delta t;r,\gamma)  = \sum_{\ell=1}^N   p_{i\ell}(t;r,\gamma) \pi_{\ell\to j}(r,\gamma).
\end{equation}
For $\Delta t$ small, we have $p_{ij}(t+\Delta t;r,\gamma)\approx p_{ij}(t;r,\gamma)+\Delta t\frac{\partial p_{ij}(t;r,\gamma)}{\partial t}$. Hence,
\begin{align}\nonumber
&\frac{\partial p_{ij}(t;r,\gamma)}{\partial t}=-\frac{p_{ij}(t;r,\gamma)-\sum\limits_{\ell=1}^N p_{i\ell}(t;r,\gamma)\pi_{\ell\to j}(r,\gamma)}{\Delta t}\\
&=
-\frac{1}{\Delta t}\sum_{\ell=1}^N\left[\delta_{\ell j}-\pi_{\ell\to j}(r,\gamma)\right]p_{i\ell}(t;r,\gamma).
\end{align}
Introducing the modified Laplacian operator $\hat{\mathcal{L}}(r,\gamma)$ with elements $\mathcal{L}_{ij}(r,\gamma)=\delta_{ij}-\pi_{i\to j}(r,\gamma)$, we have the master equation 
\begin{equation}
\frac{\partial p_{ij}(t;r,\gamma)}{\partial t}=-\frac{1}{\Delta t}\sum_{\ell=1}^N p_{i\ell}(t;r,\gamma)\mathcal{L}_{\ell j}(r,\gamma).
\end{equation}
In the following we re-define the time $t$ as $t/\Delta t$, which is equivalent to set $\Delta t=1$ above, {\it i.e.}, the hopping rate is unity.
\\[2mm]
In matrix form, $\hat{\mathcal{L}}(r,\gamma)=\mathbb{1}-\mathbf{\Pi}(r,\gamma)$ and it is straightforward to see that the matrices $\mathbf{\Pi}(r,\gamma)$ and $\hat{\mathcal{L}}(r,\gamma)$ have the same set of left and right eigenvectors $\{\left\langle\bar{\psi}_m(r,\gamma)\right|\}_{m=1}^N$ and $\{\left|\psi_m(r,\gamma)\right\rangle\}_{m=1}^N$. The  eigenvalues of  $\hat{\mathcal{L}}(r,\gamma)$ denoted as $\{\xi_m(r,\gamma) \}_{m=1}^N$, satisfy 
\begin{equation}
\xi_m(r,\gamma)=1-\zeta_m(r,\gamma) \qquad\text{for}\qquad m=1,2,\ldots,N.
\end{equation} 
Once the spectral properties of  $\hat{\mathcal{L}}(r,\gamma)$ are known, the occupation probability $p_{ij}(t;r,\gamma)$ to reach $j$ at time $t$ starting from $i$ is given by
\begin{equation}
p_{ij}(t;r,\gamma)=\sum_{\ell=1}^N \exp\left[-\xi_{\ell}t\right]\left\langle i|\psi_\ell(r,\gamma)\right\rangle \left\langle\bar{\psi}_\ell(r,\gamma)|j\right\rangle.
\end{equation} 
This expression allows us to deduce different quantities of interest for the continuous-time random walker, such as the average probability of return 
\begin{align}\nonumber
&\bar{p}_{0}(t;r,\gamma)\equiv\frac{1}{N}\sum_{i=1}^N p_{ii}(t;r,\gamma)\\\nonumber
&=\frac{1}{N}\sum_{i=1}^N\sum_{\ell=1}^N \exp\left[-\xi_{\ell}t\right]\left\langle i|\psi_\ell(r,\gamma)\right\rangle \left\langle\bar{\psi}_\ell(r,\gamma)|i\right\rangle\\\nonumber
&=\frac{1}{N}\sum_{\ell=1}^N \exp\left[-\xi_{\ell}t\right]\sum_{i=1}^N \left\langle\bar{\psi}_\ell(r,\gamma)|i\right\rangle
\left\langle i|\psi_\ell(r,\gamma)\right\rangle\\
&=\frac{1}{N}\sum_{\ell=1}^N \exp\left[-\xi_{\ell}t\right]
\end{align}
or the stationary distribution
\begin{equation}\label{EstPCont}
p_j^{\infty}(r,\gamma)\equiv\lim_{t\to \infty}p_{ij}(t;r,\gamma)=\left\langle i|\psi_1(r,\gamma)\right\rangle \left\langle\bar{\psi}_1(r,\gamma)|i\right\rangle, 
\end{equation}
which stems from the fact that $\xi_1(r,\gamma)=0$ and $\xi_m(r,\gamma)>0$ for $m=2,3,\ldots,N$. The stationary distribution in Eq. (\ref{EstPCont}) agrees with that of the discrete time dynamics.
\\[2mm]
The results presented in this section reveal the connection between discrete and continuous random walkers. Our findings for the spectral properties of the modified Laplacian $\hat{\mathcal{L}}(r,\gamma)$ coincide with the general formalism introduced by Rose {\it et al.} in \cite{Touchette_PRE2018} in the context of classical and quantum transport with resetting.
\section*{Acknowledgments}  
P. Herringer acknowledges financial support from the University of Calgary Undergraduate International Research Grant.


%

\end{document}